\documentclass[aps,prl,twocolumn,showpacs,superscriptaddress,nofootinbib,groupedaddress]{revtex4} 

\usepackage{pstricks}
\usepackage{color}
\usepackage{graphicx}
\usepackage{amsmath}
\usepackage{amssymb}
\usepackage[colorlinks=true,citecolor=darkred,urlcolor=darkred, pdfborder={0 0 0}]{hyperref}

\graphicspath{{figs/}}
\definecolor{darkred}{rgb}{0.6,0,0}
\def\gsim{\raise0.3ex\hbox{$\;>$\kern-0.75em\raise-1.1ex\hbox{$\sim\;$}}}
\def\lsim{\raise0.3ex\hbox{$\;<$\kern-0.75em\raise-1.1ex\hbox{$\sim\;$}}}
\def\beqn#1{\begin{equation}\label{#1}}
\def\eeqn{\end{equation}}
\def\beqa#1{\begin{eqnarray}\label{#1}}
\def\eeqa{\end{eqnarray}}
\def\SM{$\mathrm{SU(3)_c \otimes SU(2)_L \otimes U(1)_Y}$ }
\newcommand{\sm}{{standard model }}
\def\lfv{lepton flavour violation }
\def\lnv{lepton number violation }
\def\vev#1{\left\langle #1\right\rangle}

\newcommand{\AddrAHEP}{%
  AHEP Group, Institut de F\'{i}sica Corpuscular --
  C.S.I.C./Universitat de Val\`{e}ncia, Parc Cientific de Paterna.\\
 C/ Catedratico Jose Beltran, 2 E-46980 Paterna (Val\`{e}ncia) - SPAIN}

\begin{document}

\title{Inflation and majoron dark matter in the seesaw mechanism}

\author{Sofiane M. Boucenna}
\email{msboucenna@gmail.com}
\affiliation{ \AddrAHEP }
\author{Stefano Morisi}
\email{stefano.morisi@gmail.com}
\affiliation{ DESY, Platanenallee 6, D-15735 Zeuthen, Germany.}
\author{Qaisar Shafi} \email{shafi@bartol.udel.edu}
\affiliation{Bartol Research Institute, Department of Physics and
  Astronomy, University of Delaware, Newark, DE 19716, USA.}
\author{Jos\'e W.F. Valle}
\email{valle@ific.uv.es}
\affiliation{ \AddrAHEP }
\date{\today}
%% Keywords: majoron, dark matter, inflation, PLANCK and BICEP2
\begin{abstract}
\noindent

We propose that inflation and dark matter have a common origin,
connected to the neutrino mass generation scheme. As a model we
consider spontaneous breaking of global lepton number within the
seesaw mechanism. We show that it provides an acceptable inflationary
scenario consistent with the recent CMB B-mode observation by the
BICEP2 experiment. The scheme may also account for the baryon
asymmetry of the Universe through leptogenesis for reasonable
parameter choices.

\end{abstract}

\maketitle

\section*{Introduction}

The need to account for neutrino
mass~\cite{Maltoni:2004ei,Tortola:2012te} as well as cosmological
issues such as the explanation of dark matter~\cite{Bertone:2010},
inflation~\cite{guth1981inflationary,linde2007inflationary,Starobinsky:1980te} and the
baryon asymmetry~\cite{sakharov1967violation}
suggests that the \sm must be extended.
The recent measurent by the BICEP2 experiment of the tensor-to-scalar
ratio parameter $r=0.20^{+0.07}_{-0.05}$~\cite{Ade:2014xna} of the
primordial fluctuations of the cosmic microwave background (CMB) has
caused tremendous interest, see for instance~\cite{DiBari:2014oja} and references therein.
The possible discovery of gravity waves, if confirmed, would certainly count as one of the greatest
in cosmology.
Apart from such intrinsic significance, the measurement of nonzero $r$
implies important constraints on inflationary models of the Universe.
Here we consider the simplest type-I seesaw
scenario~\cite{minkowski:1977sc,gell-mann:1980vs,yanagida:1979,mohapatra:1980ia,Lazarides:1980nt,Schechter:1980gr}~\footnote{Note
  that in~\cite{Schechter:1980gr} this was called type~II, just the
  opposite of what has become established.}  of neutrino mass
generation in which lepton number is promoted to a spontaneously
broken symmetry, within the standard \SM gauge
framework~\cite{chikashige:1981ui, Schechter:1981cv}.
In order to consistently formulate the spontaneous violation of lepton
number within the \SM model, one requires the presence of a
lepton-number-carrying complex scalar singlet, $\sigma$, coupled to the singlet
``right-handed'' neutrinos $\nu_R$.
The real part of $\sigma$ drives inflation through a Higgs
potential~\cite{Kallosh:2007wm, Rehman:2010es,Rehman:2008qs,Okada:2014lxa,Okada:2013vxa} while the imaginary
part, which is the associated Nambu-Goldstone boson, is assumed to
pick up a mass due to the presence of small explicit soft \lnv terms
in the scalar potential, whose origin we need not specify at this
stage.  For suitable masses such a majoron can account for the dark
matter~\cite{berezinsky:1993fm}, consistent with the CMB
observations~\cite{Lattanzi:2007ux}.

We show how, for reasonable parameter choices, this simplest scenario
for neutrino masses provides an acceptable inflationary scenario.  The
scheme has also the potential to account for baryogenesis through
leptogenesis. A previous attempt relating inflation to neutrinos can
be found in~\cite{Nakayama:2013nya} where a supersymmetric model was
suggested in which the right-handed sneutrino drives chaotic
inflation.

\section*{Preliminary considerations}
\label{sec:basic-considerations}

Our model is the simplest type-I seesaw extension of the standard \SM
model with a global lepton number symmetry.  In addition to the \sm
fields we add three generations of right-handed neutrinos and a complex singlet
$\sigma$ carrying two units of lepton number.  
The relevant invariant Yukawa interactions are
\begin{equation}
  \label{eq:Yuk}
  \mathcal{L}_y = - Y _D^{ij} \overline{\ell^j_L} \imath \tau_2 \Phi^* \nu^i_R -\frac{1}{2} Y_N^{i} \sigma \,\overline{\nu_R^{ic}}\ \nu_R^i + \mathrm{h.c.}\,,
\end{equation}
where $\ell$ denotes the lepton doublet, $\Phi$ is the Higgs boson and
$\tau_2$ is the second Pauli matrix. 
After symmetry breaking characterized by the \lnv scale
$v_{\rm L}=\vev{\sigma}$~\cite{chikashige:1981ui,Schechter:1981cv} and
the usual electroweak scale $\vev{\Phi}\equiv v_2$ the resulting
seesaw scheme is characterized by singlet and doublet neutrino mass
terms, described by
\begin{equation}
\label{ss-matrix} 
{\mathcal M_\nu} = 
\begin{bmatrix}
    0 & Y _D v_2 \\[+1mm]
    {Y _D}^{T} v_2  & Y_N v_{\rm L} \\
\end{bmatrix}\, ,
\end{equation}
in the basis $\nu_{L}$, $\nu_{R}$.

The Yukawa coupling matrix $Y _D$ generates the ``Dirac'' neutrino
mass term, while $Y_N$ gives the right-handed Majorana mass term.
While the former is in principle arbitrary, the matrix $Y_N$
characterizing the coupling of $\sigma$ to the right-handed neutrinos
is symmetric and can be taken diagonal and with real positive entries
without loss of generality.
The effective light neutrino mass, obtained by perturbative
diagonalization of Eq.~(\ref{ss-matrix}) is of the form
\begin{equation}
  \label{eq:ss-formula0}
  m_{\nu} \simeq %% Y_3 v_3 -
Y _D {Y_N}^{-1} {Y _D}^T \frac{{v_2}^2}{v_{\rm L}} 
\end{equation}
This relation is consistent with tiny neutrino masses of order
$10^{-1}$ electron volt. For example, assuming $Y_D$ of
$\mathcal{O}(1)$, one needs $v_{\rm L}\gsim10^{14}$~GeV
\begin{equation}\label{YN}
Y_N\approx \frac{10^{14}\,\, {\rm GeV} }{v_{\rm L}}\,.
\end{equation}

%
%%%%%%%%%%%%%%%%%%%%%%%%%%%%%%%%%%
%%%%%%%%%%% ABOVE&BELOW for N=60
%%%%%%%%%%%%%%%%%%%%%%%%%%%%%%%%%%
%
\begin{table*}[!ht]
\centering
\begin{tabular}{|l|l|l|l|l|l|l|l|l}
% \hline
% \multicolumn{8}{c}{}\\
\multicolumn{8}{c}{Solutions above the VEV ($\rho>v_{\mathrm L}$)}\\
% \multicolumn{8}{c}{}\\
\hline
$v_L (M_P)$&$\mathrm{log}_{10}(\lambda)$&$n_s$&$r$&$\alpha\, (10^{-4})$&$V^{1/4}\, (10^{16}\,\mathrm{GeV})$&$\rho_0\, (M_P)$&$\rho_e\, (M_P)$\\
\hline
\hline
1.&-12.8521&0.951168&0.260263&-7.96468&2.30678&22.2218&3.14626\\
%2.&-12.8815&0.951978&0.25545&-7.74718&2.29604&22.562&3.8637\\
%3.&-12.9202&0.952955&0.249498&-7.50094&2.28254&23.0408&4.73084\\
%4.&-12.9637&0.953953&0.243258&-7.26609&2.26814&23.6163&5.65686\\
5.&-13.0093&0.954908&0.237136&-7.05625&2.25373&24.2634&6.61037\\
%6.&-13.0556&0.955793&0.231318&-6.87412&2.23978&24.9654&7.57863\\
%7.&-13.1018&0.9566&0.225884&-6.71807&2.22651&25.7105&8.55564\\
%8.&-13.1473&0.957331&0.220856&-6.585&2.21401&26.4901&9.53825\\
%9.&-13.1918&0.957989&0.216226&-6.47156&2.20232&27.2978&10.5246\\
% 
10&-13.2351&0.958581&0.211972&-6.37463&2.1914&28.1285&11.5137\\
20.&-13.599&0.962148&0.184081&-5.89025&2.11546&37.1396&21.4642\\
%30.&-13.8501&0.963114&0.172802&-5.92158&2.08228&46.5503&28.6191\\
%40.&-14.0576&0.963946&0.164596&-5.84548&2.05711&56.3038&38.6108\\
% 
50.&-14.2262&0.964453&0.159253&-5.80242&2.04021&66.1458&48.6058\\
100&-14.7789&0.965456&0.147557&-5.72255&2.00167&115.805&98.5958\\
%200.&-15.3572&0.965936&0.141184&-5.6854&1.9797&215.621&198.591\\
%300.&-15.7016&0.96609&0.13898&-5.67329&1.97193&315.557&298.589\\
%400.&-15.9477&0.966166&0.137864&-5.66728&1.96795&415.525&398.588\\
% 
500.&-16.1392&0.966211&0.137189&-5.66368&1.96554&515.506&498.588\\
%600.&-16.3106&0.966797&0.134495&-5.47578&1.95582&615.623&601.416\\
%700.&-16.4289&0.966262&0.136413&-5.65957&1.96276&715.484&698.587\\
%800.&-16.5586&0.966834&0.133933&-5.47286&1.95377&815.607&801.415\\
%900.&-16.6457&0.96629&0.13598&-5.65729&1.9612&915.471&898.587\\

1000.&-16.7367&0.9663&0.135828&-5.6565&1.96065&1015.47&998.587\\
\hline
% \multicolumn{8}{c}{}\\
\multicolumn{8}{c}{Solutions below the VEV ($\rho<v_{\mathrm L}$)}\\
% \multicolumn{8}{c}{}\\
\hline
$v_L (M_P)$&$\mathrm{log}_{10}(\lambda)$&$n_s$&$r$&$\alpha\, (10^{-4})$&$V^{1/4}\, (10^{16}\,\mathrm{GeV})$&$\rho_0\, (M_P)$&$\rho_e\, (M_P)$\\
\hline
\hline
%7.&-14.5649&0.836719&0.0000498536&-0.0254368&0.271379&0.0305795&5.72721\\
8.&-13.9086&0.87488&0.000385304&-0.150585&0.452484&0.111018&6.70982\\
9.&-13.5255&0.900769&0.00148882&-0.460638&0.6344&0.27599&7.69622\\
10.&-13.3033&0.918822&0.00377031&-0.949789&0.800289&0.541141&8.68529\\
15.&-13.1004&0.95579&0.0279442&-3.49461& 1.32046&3.17548&13.6523\\
20.&-13.2562&0.964198&0.0518562&-4.54129&1.54118&7.05055&18.6357\\
30.&-13.5959&0.967596&0.0798131&-5.09597&1.71661&16.0451&28.6191\\
%40.&-14.8603&0.968058&0.093905&-5.2401&1.78782&25.5942&38.6108\\
50.&-14.0675&0.96807&0.102141&-5.30133&1.8258&35.3404&48.6058\\
%100.&-14.7044&0.967683&0.11783&-5.3918&1.8922&84.8692&98.5958\\
%200.&-15.3269&0.967344&0.125203&-5.42885&1.92113&184.651&198.591\\
%300.&-15.6718&0.966663&0.12977&-5.62537&1.93841&284.707&301.418\\
%400.&-15.9253&0.966596&0.130955&-5.63136&1.94282&384.673&401.417\\
500.&-16.1213&0.966555&0.131662&-5.63496&1.94544&484.653&501.416\\
%600.&-16.2812&0.966528&0.132131&-5.63735&1.94717&584.64&601.416\\
%700.&-16.4161&0.966508&0.132465&-5.63907&1.9484&684.63&701.416\\
%800.&-16.5473&0.967048&0.130507&-5.45536&1.94116&784.494&798.587\\
%900.&-16.6358&0.966481&0.13291&-5.64134&1.95003&884.617&901.415\\
1000.&-16.7278&0.966472&0.133065&-5.64214&1.9506&984.613&1001.42\\
\hline
\end{tabular}
\caption{Higgs inflation scenario (no radiative corrections): The values of parameters for number of e-folds $N = 60$.}
\label{tab:higgs}
\end{table*}

\section*{Scalar potential}
\label{sec:scalar-potential}

We now turn to the dynamical justification of this
scenario~\footnote{ For simplicity, we take a one-generation neutrino seesaw scheme with 0.1 eV mass scale in the analysis of our proposed inflationary scenario.}, starting from the scalar potential.
The tree level Higgs potential associated with the singlet and doublet
scalar multiplets $\sigma$ and $\Phi$ is a simple extension of that
which characterizes the standard model,
\begin{equation}
\label{eq:Vtree}
 V_{\mathrm{tree}} = \lambda \left( 
  \sigma^\dagger \sigma - \frac{v_{\mathrm{L}^2}}{2} \right)^2 + \lambda_{\mathrm{mix}} (\sigma^\dagger \sigma) (\Phi^\dagger \Phi)  + V_\Phi,   
\end{equation}
where $V_\Phi$ is the SM potential.  As will become clear later,
inflation and neutrino masses require that $\vev{\sigma}\gg
\vev{\Phi}$.  We also consider $\lambda_{\mathrm{mix}}$ to be
negligible in order to use the small decay width
approximation~\cite{Okada:2013vxa}.  
The inflaton is identified with the real part of $\sigma$ 
\begin{equation}
  \label{eq:inflaton}
  \rho \equiv \sqrt{2}~\Re[\sigma]   \, ,
\end{equation}
and we parametrize the effective potential in the leading-log
approximation, with the renormalization scale fixed at
$v_{\mathrm{L}}$, as~\cite{coleman:1973jx}
\begin{equation}
\label{eq:effpot}
V = \lambda
\left[ 	
\frac{1}{4}\left( 
\rho^2 - v_{\mathrm{L}}^2
\right)^2 + 
a \log \left[\frac{\rho}{v_{\mathrm{L}}} \right] \rho^4
+V_0 \right], 
\end{equation}
where $a= \frac{\beta_\lambda}{16 \pi^2 \lambda}$ and the coefficient 
$\beta_\lambda$ is given as
\begin{eqnarray}\label{eq:beta}
\beta_\lambda &=&
20 \lambda^2 +  2 \lambda\left( \sum_i (Y_N^i)^2 \right) - \sum_i (Y_N^i)^4. \nonumber \\
&\simeq&  - \sum_i (Y_N^i)^4.
\end{eqnarray}
The last approximation $\lambda\ll Y_N$ will be justified later. An
analysis of the potential reveals that $a\gsim -0.2$ ensures a
consistent local minimum.

\section*{Inflation scenarios}

\begin{figure}[t!]
\begin{center}
\includegraphics[width=0.45\textwidth]{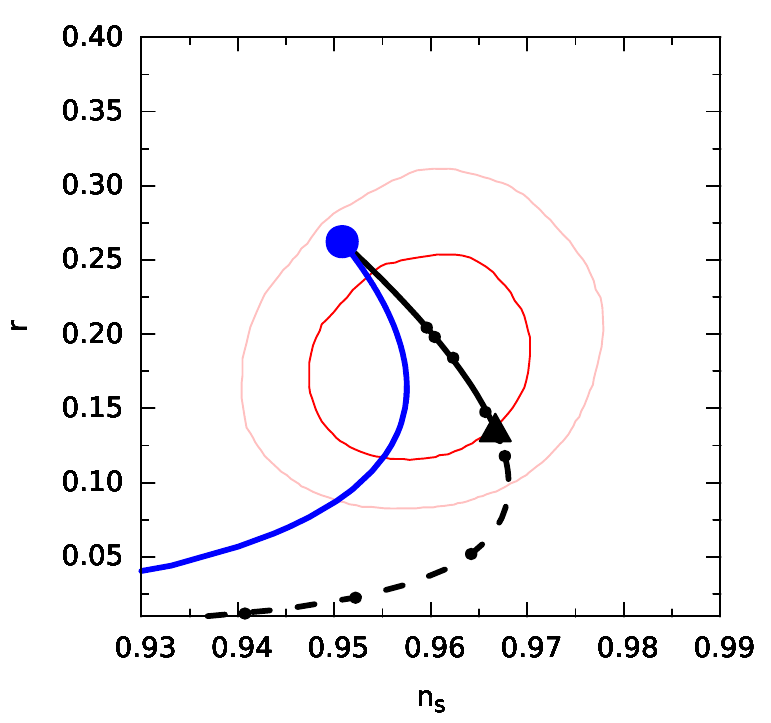}
\caption{Majoron Inflation: The tensor-to-scalar ratio $r$ is shown versus 
  the spectral index $n_s$. Black line is the Majoron inflation scenario with 
  $v_{\rm L} > M_P$. The small black points on each branch, from left
  to right, indicate the values $v_{\rm L}/M_P=12,14,20$ and $100$.  The dashed branch corresponds to $\sigma<v_{\rm L}$ and
  the solid one to $\sigma>v_{\rm L}$. The point and the triangle are 
  the quartic and quadratic inflation predictions, respectively.  
  The blue (gray) line is for $v_{\rm L} \ll M_P$.  The contours
  are the 68\% and 95\% CL allowed region, combining PLANCK, WP, highL and BICEP2, given in~\cite{Ade:2014xna} 
  and $N$ is taken to be 60.}
 \label{rns}
\end{center}
\end{figure}

Here we consider the radiatively corrected $\rho^4$ potential.
Inflation takes places as the inflaton slowly rolls down to the
potential minimum either from above ($\sigma>v_{\rm L}$) or from
below ($\sigma<v_{\rm L}$).
The inflationary slow-roll parameters are given by
\begin{eqnarray}
\label{eq:inflation-slow-roll}
 \epsilon (\rho) &=& \frac{1}{2} M_P^2 
 \left( \frac{V^\prime}{V} \right)^2, 
\; \;  \eta (\rho) = M_P^2 \left( \frac{V''}{V} \right), \nonumber \\ 
\zeta^2 (\rho) &=& M_P^4 \left(
 \frac{V'  V'''}{V^2} \right), 
\end{eqnarray}
where prime denotes a derivative with respect to $\rho$ and
$M_P=2.4\times10^{18}$ is the (reduced) Planck mass.  The slow-roll
approximation is valid as long as the conditions $\epsilon, |\eta|,
\zeta^2 \ll 1$ hold.  In this case, the scalar spectral index $n_{s}$,
the tensor-to-scalar ratio $r$, and the running of the spectral index
$\alpha$ are given by
\begin{eqnarray}\label{eq:slowroll-index}
 n_s &\simeq& 1-6 \epsilon + 2 \,,\quad  r \simeq 16 \epsilon \,,  \nonumber \\ 
\alpha &\equiv&  \frac{d n_{s}}{d \ln k} \simeq 16 \epsilon  \eta 
   - 24  \epsilon^2 - 2 \zeta^2. 
\end{eqnarray}
The amplitude of the curvature perturbation $\Delta_{\mathcal{R}}$ is
\begin{equation}
\label{eq:Delta-inflation}
 \Delta_{\mathcal{R}}^2 = \left. \frac{V}{24\,\pi^2 \,
M_P^4\,\epsilon } 
 \right|_{k_0},
\end{equation}
and is taken as $ \Delta_{\mathcal{R}}^2=2.215\times 10^{-9}$ to fit
PLANCK CMB anisotropy measurements~\cite{Ade:2013ktc}, with the pivot
scale chosen at $k_0 = 0.05\, {\rm Mpc^{-1}}$. Finally, the number of
e-folds realized during inflation is
\begin{equation}
\label{eq:efolds-slow-roll}
N= \frac{1}{\sqrt{2} M_P} 
 \int_{\rho_e}^{\rho_0}
 \frac{d \rho}{\sqrt{\epsilon(\rho)}}\hspace{0.3cm},
\end{equation}
where $\rho_0$ is the field value that corresponds to $k_0$ and
$\rho_e$ denotes the value of $\rho$ at the end of inflation, ie. when
$\epsilon(\rho_e) \approx 1$.

At this stage we have four parameters ($Y_D$, $a$, $v_{\rm L}$ and
$\lambda$) for five observables ($m_\nu$, $r$, $n_s$, $\alpha$ and
$\Delta_{\mathcal{R}}^2$).  Once we calculate $\rho_e$ and $\rho_0$,
$\lambda$ is fixed from the constrain on $\Delta_{\mathcal{R}}^2$ and
we find that $\lambda\approx 10^{-17}-10^{-12}$ in the parameter space of the
model, which justifies the approximation made in Eq.~(\ref{eq:beta}). We
are then left with $a$ (ie. $Y_N$), $Y_D$ and $v_{\rm L}$ and neutrino
masses further constrain the relation between $Y_N$ and $Y_D$. The
predicted values of $r$, $n_s$ and $\alpha$ are therefore predicted
for fixed values of $a$ and $v_{\rm BL}$. 

We will consider two limits: $v_{\rm L}> M_P$, the so-called Higgs
inflation as well as $v_{\rm L}\ll M_P$ when the scalar potential
considered in Eq.~(\ref{eq:effpot}) reduces to the radiatively corrected
quartic inflation~\cite{NeferSenoguz:2008nn}.

\subsubsection{I. Higgs inflation}
\begin{figure}[h!]
\begin{center}
\includegraphics[width=0.45\textwidth]{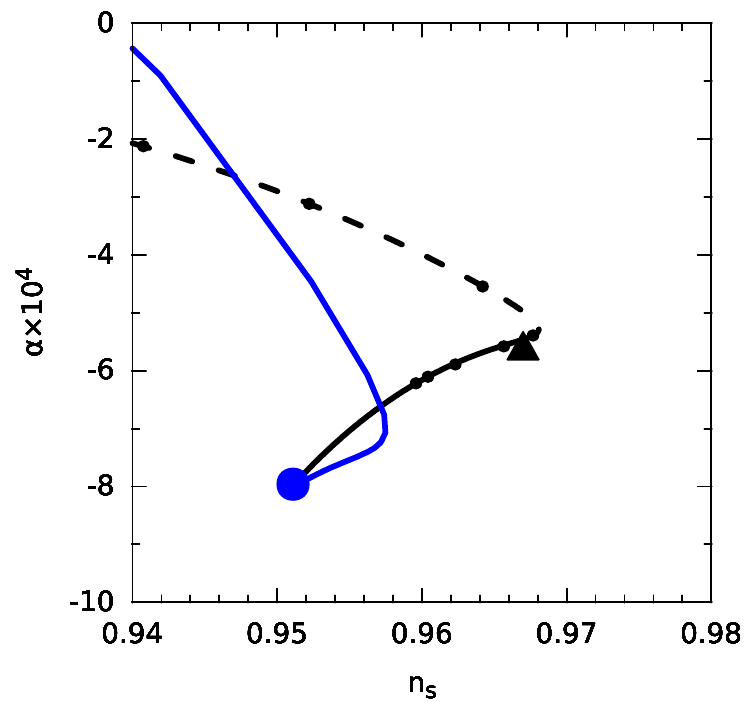}
\caption{Majoron inflation: $\alpha$ vs. $n_s$ for various $v_{\rm
    BL}$ values. See caption of Fig.~(\ref{rns}) for more details.}
\label{nsalpha}
\end{center}
\end{figure}

\begin{figure}[h!]
\begin{center}
\includegraphics[width=0.45\textwidth]{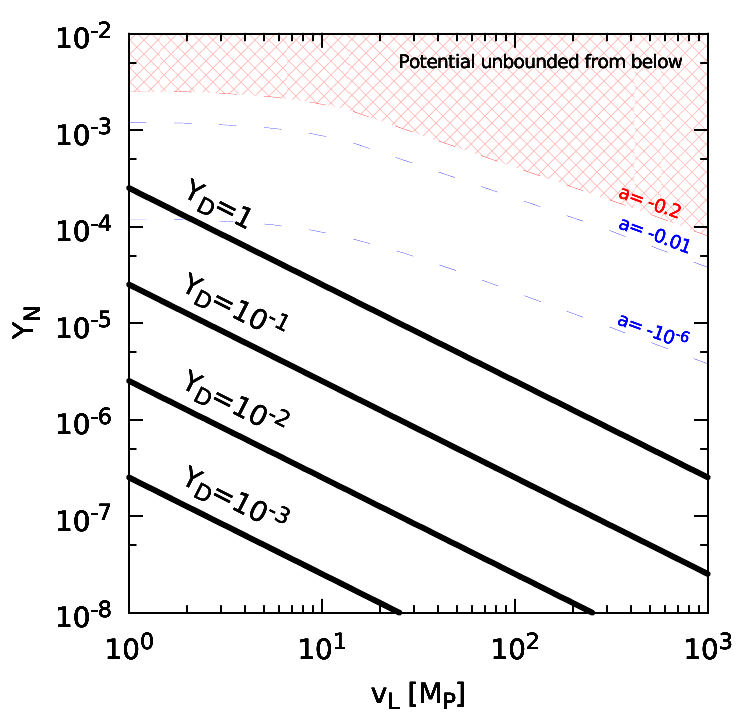}
\caption{Majoron inflation: $Y_N$ vs. $v_{\rm L}$ for various
  $Y_D$. Dashed lines show some values of the coefficient $a$ of the
  Coleman-Weinberg term in the potential. Solid black lines are upper
  bounds on $Y_N$ for the corresponding Dirac neutrino Yukawa coupling
  $Y_D$. }
\label{ynvbl}
\end{center}
\end{figure}

This scenario requires trans-Planckian vevs. The seesaw relation,
Eq.~(\ref{YN}) imposes $Y_N\ll 1$ in order to suppress the right handed
neutrino mass.  For instance for $v_{\rm L} = 10^3\, {\rm M_P}$, one
gets $Y_N\approx 10^{-6}$, a value similar to the electron Yukawa
coupling.  The Coleman-Weinberg radiative corrections are negligible
in this case and we consider only the tree level potential.  Black
lines in Fig.~(\ref{rns}) show the predicted values of $r$ and $n_s$
obtained by varying $v_{\rm L}$ and taking the number of e-foldings
$N=60$.  The allowed 68\% and 95\% CL contours are indicated. The
dashed line is when the inflaton rolls from ``below" ($\rho<v_{\rm
  BL}$) while the solid one is for the opposite case. Both branches
converge toward quadratic (indicated by a triangle) inflation in the
limit $\rho\to\infty$, ($n_s$, $r$)=(0.967, 0.132). We show various
values of $v_{\rm L}$ as small circles. The small vev limit, depicted
by a big circle corresponds to the textbook quartic inflation
potential, ($n_s$, $r$)=(0.951, 0.262).
The running of the spectral index, $\alpha$, is depicted in
Fig.~(\ref{nsalpha}).
In Fig.~(\ref{ynvbl}) we show the connection between inflation and
neutrino masses, in the plane $Y_N$ vs. $v_{\rm L}$. The black lines
are upper bounds on $Y_N$ for a given Dirac coupling $Y_D$. We also
show some values of $a$ corresponding to each $Y_N$ and $v_{\rm L}$
for completeness. The numerical results for this case are displayed in Tab. (\ref{tab:higgs}).
\subsubsection{II. Quartic inflation}
The sub-Planckian inflationary scenario $v_{\rm L} \ll M_P$, in
principle physically more attractive, is well approximated by the
quartic potential.  In this case, $Y_N$ can be large so that the
radiative corrections to the $\rho^4$ potential should be taken into
account.  The quantum corrections allow us to depart from the fixed
textbook prediction of quartic inflation to lie closer to the BICEP2
region.  Fig.~(\ref{rns}) and Fig.~(\ref{nsalpha}) show the effect of the
coupling of the inflaton to right handed neutrinos on the inflationary
observables.  The blue line, departing from the quartic inflation
prediction is obtained by varying $a$, and consequently $Y_N$ in the
range [-0.2, 0] corresponding to a variation of $Y_N$ around $\approx
10^{-3}$. If $v_{\rm L}$ is taken to lie around $10^{14}\, {\rm GeV}$
then $Y_N\approx 10^{-2}$ reproduces the correct neutrino mass scale.
We display in Tab. (\ref{tab:rad}) the numerical results for this case.
%%
%%%%%%%%%%%%%%%%%%%%%%%%%%%%%%%%%%
%%%%%%%%%%% Radiative phi4 - N=60.
%%%%%%%%%%%%%%%%%%%%%%%%%%%%%%%%%%
%
\begin{table*}[!ht]
\centering
\begin{tabular}{|l|l|l|l|l|l|l|l|l|l}
% \hline
% \multicolumn{8}{c}{}\\
\multicolumn{9}{c}{Small solutions ($0.01 \lesssim r \lesssim 0.02$)}\\
% \multicolumn{8}{c}{}\\
\hline
$a$&$|Y_N|$&$\mathrm{log}_{10}(|\lambda|)$&$n_s$&$r$&$\alpha\, (10^{-4})$&$V^{1/4}\, (10^{16}\,\mathrm{GeV})$&$\rho_0\, (M_P)$&$\rho_e\, (M_P)$\\
\hline
\hline
%-0.01287&0.00172282&-11.363&0.946105&0.0724935&-2.23833&1.67582&17.719&2.53111\\
%-0.01286&0.00172599&-11.3595&0.947267&0.0762631&-2.63035&1.69719&17.8101&2.5329\\
%-0.01285&0.00172829&-11.3569&0.948328&0.0800019&-2.9958&1.71762&17.8982&2.53467\\
-0.01307&0.00135604&-11.7856&0.890248&0.0100493&7.9222&1.02256&15.1923&2.49121\\
-0.01305&0.00142537&-11.6983&0.899145&0.0137211&7.32328&1.10535&15.5053&2.49559\\
-0.01304&0.00145721&-11.6596&0.903321&0.0158434&6.92563&1.14582&15.6575&2.49774\\
-0.01303&0.00148709&-11.624&0.907307&0.0181559&6.47185&1.18552&15.8065&2.49987\\
-0.01302&0.00151498&-11.5914&0.911098&0.0206547&5.97014&1.22435&15.9522&2.50198\\
\hline
% \multicolumn{8}{c}{}\\
\multicolumn{9}{c}{Large solutions ($0.1 \lesssim r \lesssim 0.2$)}\\
% \multicolumn{8}{c}{}\\
\hline
$a$&$|Y_N|$&$\mathrm{log}_{10}(|\lambda|)$&$n_s$&$r$&$\alpha\, (10^{-4})$&$V^{1/4}\, (10^{16}\,\mathrm{GeV})$&$\rho_0\, (M_P)$&$\rho_e\, (M_P)$\\
\hline
\hline
-0.01279&0.00172752&-11.3556&0.952953&0.101404&-4.68889&1.82249&18.3706&2.54494\\
%-0.01278&0.00172549&-11.3573&0.953491&0.104754&-4.89947&1.83736&18.4408&2.54659\\
%-0.01277&0.00172306&-11.3594&0.953976&0.108033&-5.09307&1.85158&18.5088&2.54823\\
%-0.01276&0.00172025&-11.3619&0.954414&0.111241&-5.27093&1.86517&18.5746&2.54986\\
%-0.01275&0.00171711&-11.3647&0.954807&0.114375&-5.43425&1.87817&18.6384&2.55146\\
%-0.01274&0.00171368&-11.3679&0.95516&0.117435&-5.58416&1.89061&18.7003&2.55306\\
%-0.01273&0.00170998&-11.3713&0.955477&0.120422&-5.72174&1.90252&18.7602&2.55464\\
%-0.01272&0.00170604&-11.375&0.95576&0.123334&-5.84798&1.91392&18.8183&2.5562\\
%-0.01271&0.00170189&-11.3788&0.956012&0.126173&-5.96382&1.92484&18.8746&2.55775\\
%-0.0127&0.00169755&-11.3829&0.956236&0.128938&-6.07015&1.9353&18.9293&2.55928\\
%-0.01269&0.00169305&-11.3872&0.956435&0.131632&-6.16776&1.94533&18.9823&2.5608\\
%-0.01268&0.00168841&-11.3916&0.956611&0.134254&-6.2574&1.95494&19.0338&2.56231\\
%-0.01267&0.00168364&-11.3962&0.956766&0.136806&-6.33975&1.96417&19.0838&2.5638\\
%-0.01266&0.00167876&-11.4009&0.956901&0.13929&-6.41546&1.97302&19.1324&2.56528\\
-0.01265&0.00167379&-11.4057&0.957019&0.141706&-6.48511&1.98152&19.1795&2.56674\\
%-0.01264&0.00166873&-11.4106&0.95712&0.144056&-6.54921&1.98969&19.2254&2.56819\\
%-0.01263&0.00166361&-11.4156&0.957207&0.146342&-6.60826&1.99754&19.2699&2.56963\\
%-0.01262&0.00165844&-11.4207&0.957281&0.148565&-6.6627&2.00508&19.3132&2.57106\\
-0.01261&0.00165322&-11.4258&0.957343&0.150727&-6.71294&2.01234&19.3554&2.57247\\
%-0.0126&0.00164796&-11.431&0.957394&0.15283&-6.75934&2.01932&19.3964&2.57387\\
%-0.01259&0.00164268&-11.4362&0.957435&0.154874&-6.80225&2.02604&19.4363&2.57525\\
%-0.01258&0.00163738&-11.4415&0.957467&0.156863&-6.84196&2.03251&19.4751&2.57663\\
%-0.01257&0.00163206&-11.4468&0.957491&0.158797&-6.87876&2.03874&19.5129&2.57799\\
-0.01256&0.00162674&-11.4521&0.957507&0.160678&-6.9129&2.04476&19.5497&2.57934\\
%-0.01255&0.00162142&-11.4575&0.957517&0.162507&-6.94461&2.05055&19.5856&2.58068\\
%-0.01254&0.0016161&-11.4628&0.957521&0.164287&-6.97411&2.05614&19.6205&2.58201\\
%-0.01253&0.00161079&-11.4682&0.957518&0.166018&-7.00157&2.06154&19.6546&2.58332\\
%-0.01252&0.00160549&-11.4736&0.957511&0.167703&-7.02718&2.06675&19.6879&2.58463\\
%-0.01251&0.00160021&-11.479&0.9575&0.169342&-7.0511&2.07178&19.7203&2.58592\\
-0.0125&0.00159495&-11.4843&0.957484&0.170937&-7.07347&2.07664&19.7519&2.5872\\
-0.0124&0.00154397&-11.5373&0.957174&0.184759&-7.2355&2.1174&20.0299&2.59943\\
-0.0123&0.00149676&-11.5877&0.956735&0.195481&-7.33264&2.14748&20.2527&2.61069\\
-0.0122&0.00145363&-11.635&0.956276&0.20395&-7.39978&2.17037&20.4349&2.62107\\
-0.0121&0.00141436&-11.679&0.95584&0.210759&-7.45154&2.18826&20.5865&2.63068\\
-0.0119&0.00134587&-11.7579&0.955081&0.220938&-7.53147&2.21422&20.8243&2.64788\\
-0.0116&0.00126256&-11.8579&0.954217&0.230944&-7.62064&2.23887&21.0753&2.66959\\
\hline
\end{tabular}
\caption{Radiatively corrected quartic potential: The values of parameters for number of e-folds $N = 60$}
\label{tab:rad}

\end{table*}
\section*{Dark matter and leptogenesis}
In the limit where lepton number is an exact symmetry of the
Lagrangian, \lnv is purely spontaneous so that the associated
Nambu-Goldstone boson, i.e. the majoron, given as the imaginary part
of $\sigma$, is strictly massless. However soft explicit \lnv may arise
from a variety of sources, including quantum gravity
effects~\cite{kallosh:1995hi,Akhmedov:1992hi}. Motivated by these considerations in
fact the KeV majoron has been suggested as a viable dark matter
candidate~\cite{berezinsky:1993fm} much before the precise CMB
observations from WMAP and PLANCK were available.
Being a Goldstone boson associated with the spontaneous breaking of
ungauged lepton number, the massive majoron will decay to a pair of
neutrinos through a small coupling dictated by Noether's theorem to be
proportional to the small neutrino mass~\cite{Schechter:1981cv}. 
The existence of this two--neutrino decay mode modifies the power
spectrum of the cosmic microwave background temperature
anisotropies~\cite{Lattanzi:2007ux}. One can determine the majoron
lifetime and mass values required by the CMB observations in order for
the majoron dark matter picture of the Universe to be consistent.
It has been shown that the majoron provides an acceptable
decaying dark matter scenario for suitably chosen mass
values~\cite{Lattanzi:2013uza} which depend on whether or not the
majorons are thermal or not.
If the majoron production cannot be thermal, as it may be the case in
the first inflationary scenario we considered, due to the smallness of
the $Y_N$ and $\lambda_{mix}$ couplings, one can still consider
non-thermal mechanisms such as freeze-in~\cite{Frigerio:2011in} or
scalar field oscillations~\cite{Turner:1983he,Kazanas:2004kv}.
Moreover, in such non-thermal case, the mass of the majoron is not
constrained to be of $\mathcal{O}$(KeV) and can lie in a large range
depending on the details of the mechanism under consideration.

Turning now to leptogenesis~\cite{Fukugita:1986hr} we note that after
spontaneous \lnv occurs at the scale $v_{\rm L}$ the type I seesaw
mechanism is generated and the Universe reheats at the same time.
The presence of right-handed neutrinos with direct couplings to the
inflaton field is an important ingredient for leptogenesis~\cite{Lazarides:1991wu}.
\section*{\bf Conclusions}
We have suggested that neutrino masses, inflation and dark matter may
have a common origin. We have illustrated this with the simplest
type-I seesaw model with spontaneous breaking of global lepton number.
The resulting inflationary scenario is consistent with the recent CMB
B-mode observation by the BICEP2 experiment. On the other hand, the
scheme may also account for majoron dark matter and possibly also
leptogenesis induced through the out-of-equilibrium decays of the
right-handed neutrinos, for reasonable parameter values. If
supersymmetry is invoked, then one has a majoron version of the
supersymmetric type I seesaw, in which \lfv processes may be
within the reach of future experiments~.\\

\section*{\bf Acknowledgments}
The work of SB and JV was supported by MINECO grants FPA2011-22975 and Multidark
Consolider CSD2009-00064. S.M. thanks DFG grant WI 2639/4-1.
Q.S. acknowledges support provided by the DOE grant No. DE-FG02-12ER41808 and
thanks Jose Valle and members of the particle theory group for their hospitality.

\bibliographystyle{h-physrev3.bst} 
\bibliography{merged,newrefs,refs}
\end{document}